\documentclass[prl,twocolumn,superscriptaddress]{revtex4}

\usepackage{graphicx,epsfig}
\usepackage{bm}
\newcommand{\CZ}{\textit{CZ}}

\begin{document}
\title{Quantum Information Processing with Delocalized Qubits under Global Control}

\author{Joseph Fitzsimons}
\email{joe.fitzsimons@materials.ox.ac.uk}
\affiliation{Department of Materials, Oxford University, Oxford OX1 3PH, UK}
\author{Li Xiao}
\affiliation{Clarendon Laboratory, Department of Physics, Oxford University, Oxford OX1 3PU, UK}
\author{Simon C. Benjamin}
\affiliation{Department of Materials, Oxford University, Oxford OX1 3PH, UK}
\author{Jonathan A. Jones}
\affiliation{Clarendon Laboratory, Department of Physics, Oxford University, Oxford OX1 3PU, UK}
\date{\today}

\maketitle

{\bf Any technology for quantum information processing (QIP) \cite{bennett00} must embody within it quantum bits (qubits) and maintain control of their key quantum
properties of superposition and entanglement. Typical QIP schemes \cite{Kane98} envisage an array of physical systems, such as electrons or nuclei, with each system
representing a given qubit. For adequate control, systems must be distinguishable either by physical separation or unique frequencies, and their mutual interactions
must be individually manipulable. These difficult requirements exclude many nanoscale technologies where systems are densely packed and continuously interacting. Here
we demonstrate a new paradigm: restricting ourselves to global control pulses~\cite{lloyd93} we permit systems to interact freely and continuously, with the consequence
that qubits can become delocalized over the entire device~\cite{raussendorf,twamley}. We realize this using NMR studies of three $^{\text{13}}\text{C}$ nuclei in
alanine, demonstrating all the key aspects including a quantum mirror, one- and two-qubit gates, permutation of densely packed qubits and Deutsch algorithms. }
\bigskip

\begin{figure}
\centering
\includegraphics[width=6.9cm]{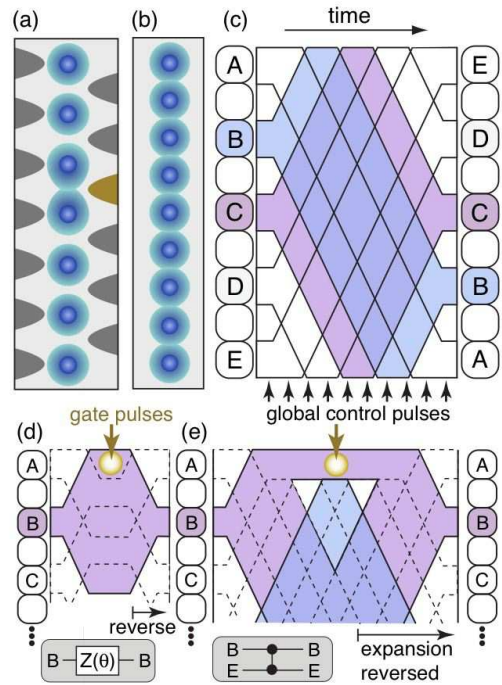}
\caption{(a) In a conventional QIP architecture qubits are localised on specific physical systems. Control over individual qubits and their interactions must be
maintained, for example by an array of adjacent electrodes. (b) Recent proposals \cite{twamley, raussendorf} show that one can instead employ an array of identical
systems, which continuously interact with their neighbors. (c) Qubits, initially localized on specific sites (left), will rapidly delocalise and overlap. Using global
control signals one can direct this process so that qubits successively delocalise, reflect from each end of the chain, and revive on specific sites. (d) In one
approach \cite{twamley}, we can exploit the {\em edges} of the array to perform operations on specific qubits as they reflect from it. The pattern of global pulses may
optionally be reversed to quickly relocalize the qubits. (e) Gate operations between qubits involve trapping part of one qubit at the edge until the second reaches it
and an interaction can be synthesized. }\label{fig:overview}
\end{figure}

Many theoretical schemes for implementing quantum information processing have been suggested. The majority of these proposals, including most solid state schemes,
involve an array of elementary systems (Fig.~\ref{fig:overview}), with some form of physical interaction coupling neighboring elements \cite{Kane98, Imamoglu, Mozyrsky,
Zagoskin}. Conventionally one aims to suppress this interaction most of the time in order to isolate qubits, thus ensuring that they do not suffer unwanted entanglement
with one another. When two qubits {\em are} required to participate in a gate operation, their interaction is `switched on' for example by altering a nearby electrode
potential~\cite{Kane98, Mozyrsky}, by moving the relevant systems into closer proximity~\cite{atomchip}, or by making a local measurement on the two
systems~\cite{loss}. However, regardless of the mechanism for this control, it constitutes a major design challenge and necessarily limits the range of systems that can
be supported. Moreover the proximity of control elements is liable to be a primary decoherence source.

The information processing potential of an array of {\em identical, permanently coupled} systems was first appreciated in the context of state transfer
\cite{Christandl04, Albanese04, Karbach05, Clark04, Yung04, Shi04}. A chain of spins with suitably engineered couplings has the property that a qubit placed on one end
will later manifest at the other, even though at intervening times it is distributed over the chain. When more than one qubit is placed on the chain, each will manifest
at the complementary site, but typically the qubits will have acquired entangling phases. This can in principle be employed to process information \cite{Clark04,
yung06} rather than simply transmitting it, but in order for a spin chain to exhibit this powerful natural dynamics it is necessary to engineer the chain for a specific
pattern of spin--spin coupling strengths \cite{yungCriterion}.

\begin{figure}
\centering
\includegraphics[width=8.2cm]{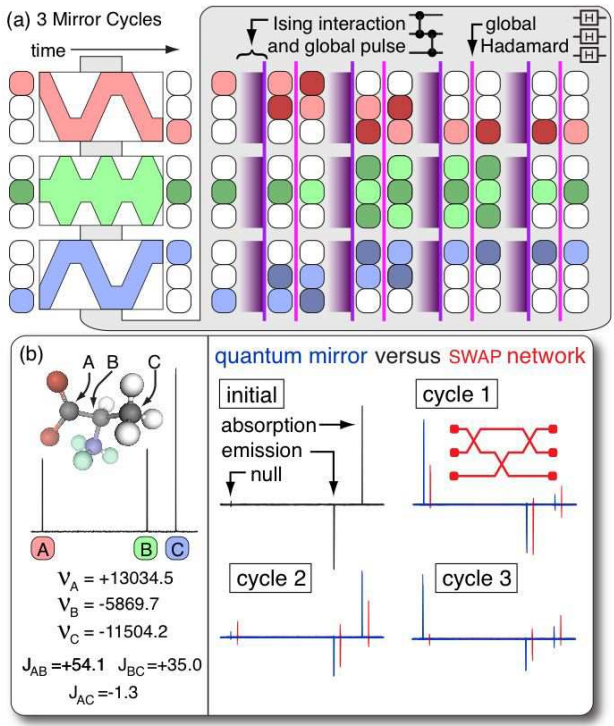}
\caption{Globally controlled quantum mirror. (a) Left panel shows the qubit `delocalise-and-revive' patterns over three mirror cycles for a three spin chain.  Right
panel shows the detailed evolution during one mirror cycle.  (b) Experimental data. Left panel shows the structure and NMR spectrum of ${}^{13}\text{C}$-labeled alanine in
$\text{D}_2\text{O}$ together with the relevant NMR parameters.  All frequencies are in Hz; following NMR conventions frequencies are measured from the rf reference
frequency and increase from right to left. Since the mirroring process should act equivalently on any initial state, we used the mixed state
$\textbf{1}+p(\sigma_z^A-\sigma_x^B+\sigma_x^C)$ with $p\sim10^{-5}$. Right panel shows spectra after successive mirroring operations and compared with conventional
\textsc{swap} networks. The initial state shows emission from spin B and absorption from spin C, with no signal from A. Successive mirroring operations exhibit
precisely the predicted behavior: states $\sigma_x$ and $\sigma_z$ exchange places, while $-\sigma_x$ delocalises over all spins and revives on B.  Note that it is not
simple to directly compare intensities on spins A and C.} \label{fig:mirroringExp}
\end{figure}

Fortunately, two recent proposals~\cite{twamley,raussendorf} have demonstrated that mirror inversion can be achieved with a {\em regular} spin chain if a global signal
pulse is applied to the system repeatedly during its dynamical evolution. In Ref.~\cite{twamley} a procedure is described whereby a spin chain with Hamiltonian
$\mathcal{H}_{\rm Ising}=J\sum_{j=1}^{N-1}\sigma_z^j\sigma_z^{j+1}$ is subjected to stroboscopic global pulses in order to synthesize a periodic update of $\CZ$ (a
controlled-phase gate between all nearest neighbors) followed by $H$ (a Hadamard gate applied to each spin). The procedure exploits the ends of a chain: the terminating
spins have a unique environment and therefore are {\em effectively} a distinct species. The global pulse prescription is the same regardless of the number of internal
spins, and will generate a perfect mirror.

In our first experimental demonstration, we have realized precisely the predicted mirror behavior in a three-spin NMR device \cite{Cory1997,Jones1998,Vandersypen2004},
specifically ${}^{13}\text{C}$-labeled alanine in solution in $\text{D}_2\text{O}$ \cite{Fortunato2002}, see Fig.~(\ref{fig:mirroringExp}). Three mirror cycles where
implemented, without loss of qubit integrity.  All gates were applied as global pulses using strongly modulated composite pulses \cite{Fortunato2002}, except that
selective pulses were applied to the end spins where necessary.  The signal loss observed arises from a combination of decoherence and accumulated errors, and the
global quantum mirror is clearly far superior to a conventional approach based on \textsc{swap} gates. During each mirror cycle, the central qubit becomes delocalized over
the whole chain and then revives, while the outer qubits delocalize over two adjacent sites. We emphasize that this experiment represents the first
successful demonstration of a fully scalable quantum mirroring procedure.

The authors of Refs.~\cite{raussendorf,twamley} proceed to show, by two quite different approaches, that the cycle of qubit delocalization and revival can be further
refined in order to perform quantum logic and therefore general quantum computation. These elegant schemes constitute perhaps the simplest models, in terms of
experimental requirements, that have ever been derived for QIP. Here we develop the scheme of Ref.~\cite{twamley}, which has the merit of explicit support for real
Ising interactions.

\begin{figure}[t]
\centering
\includegraphics[width=8cm]{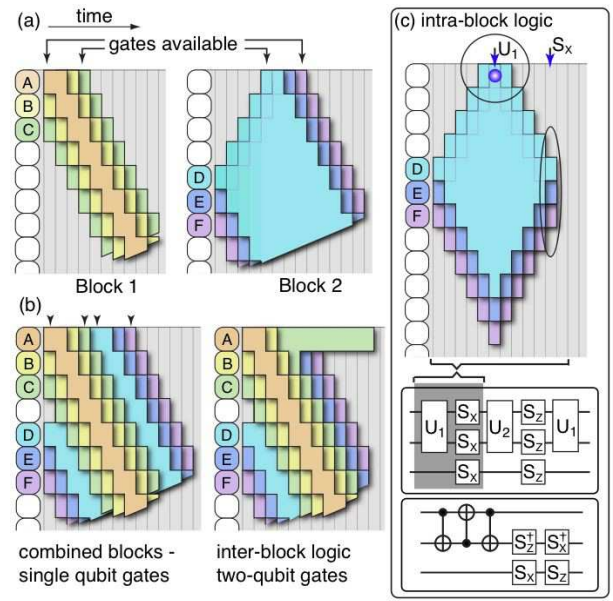}
\caption{An extension of Ref.~\cite{twamley} with a higher storage density. Qubits are stored in blocks of length $M$ along the array, with `buffer' spins marking each
division; here we depict $M=3$. In (a) two blocks are drawn separately for clarity; the system with both blocks present is drawn in (b).  The necessary {\em
intra}-block logic operations are depicted in panel (c). }\label{fig:newScheme}
\end{figure}

For state mirroring it is possible to store one qubit on each physical spin, but generally this is not true for full QIP under global control.  In previous schemes the number
of qubits is  a fraction of the total number of physical spins; according to the details of the protocol, this storage density may be an eighth~\cite{benj1}, a
quarter~\cite{raussendorf}, a third~\cite{lloyd93} or a half~\cite{benj2}. In Ref.~\cite{twamley}, the storage density has the relatively high value of one half: every
second spin is a buffer, providing separation between qubits so that, at certain moments, only one qubit `touches' the chain end and logic operations can be
synthesized, see Fig~\ref{fig:overview}.

\begin{figure}
\includegraphics[width=7cm]{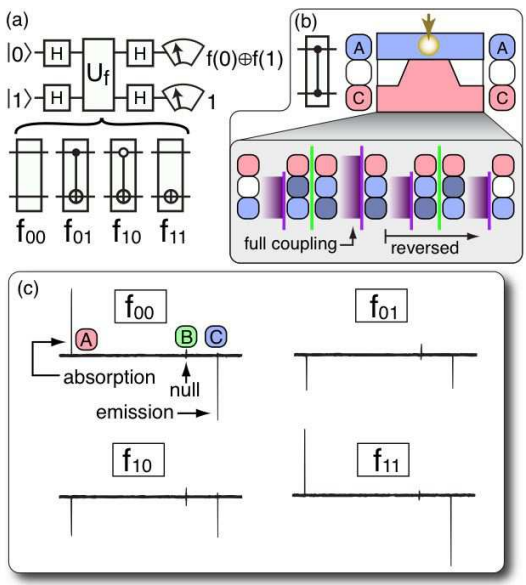}
\caption{An NMR implementation of inter-block logic operations, performing Deutsch's algorithm on a two-qubit computer with the central spin used as a buffer.  The
circuit for this algorithm is shown in (a) while (b) depicts the key two-qubit gate needed for $f_{01}$ and $f_{10}$: spin A is inverted during $\CZ$ operations
isolating its qubit from the mirror process and trapping it at the edge until the second qubit is brought into contact. Experimental results are shown in (c); as expected the signal
from spin C is in emission, spin B gives no signal, and spin A is in absorption for the constant functions $f_{00}$ and $f_{11}$ and emission for the balanced functions
$f_{01}$ and $f_{10}$.}\label{fig:interblockExps}
\end{figure}

Here we propose and demonstrate a new method of {\em block encoding} qubits so that an arbitrarily small fraction of ancillary qubits can suffice, and the storage
density approaches unity. There is an associated time overhead for storage densities above two-thirds. Our scheme, which we describe in Fig.~\ref{fig:newScheme} with
further details in the Methods and Supporting Material, involves two types of process: inter-block logic and intra-block logic.  During the cycle of delocalization and
revival the qubit at each end of every block will at some point be the sole qubit affecting the edge spin. At such times one can apply single qubit gates to these
qubits, or `freeze' them for a subsequent two-qubit gate \cite{twamley}. However, qubits {\em within} each dense block cannot be manipulated in this way. It is
therefore necessary to {\em permute} qubits within blocks in order to bring selected qubits to the border.  Gate operations $U_1=\exp(-i\pi/4\,\sigma_z^1\sigma_x^2)$,
$U_2=\exp(-i\pi/4\,\sigma_x^1\sigma_z^2)$, $S_x=\exp(-i\pi/4\,\sigma_x)$ and $S_z=\exp(-i\pi/4\,\sigma_z)$ are combined as shown to produce a qubit swap plus spurious
local gates which can be corrected. All blocks can be permuted in this way, leading to a modest time overhead factor of order $M$.

\begin{figure}
\includegraphics[width=6cm]{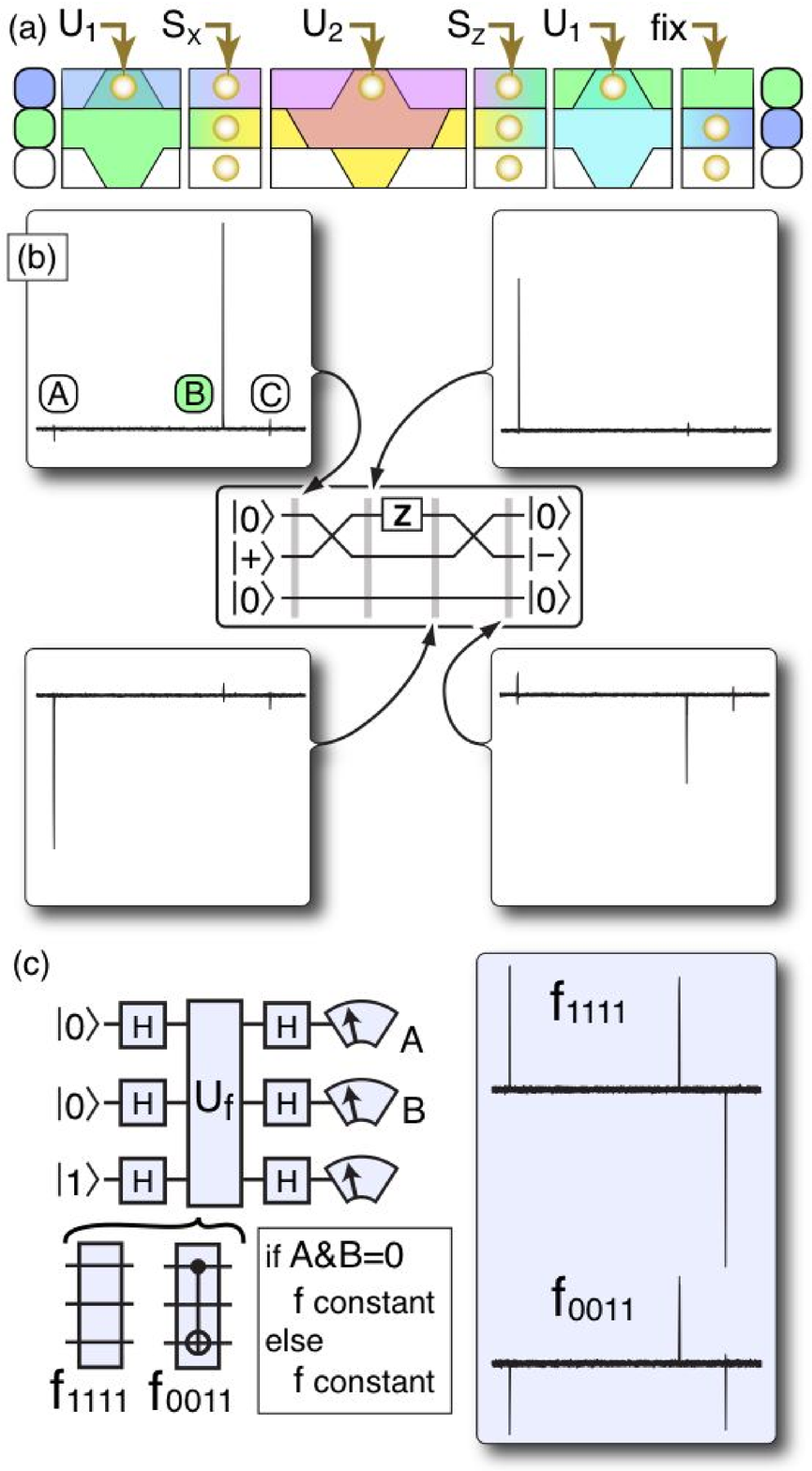}
\caption{An NMR implementation of intra-block logic on a 3 spin system.  Spectra in (b) show how a $z$-rotation can be performed on the central spin using global
control methods as shown in (a); note that the final step (fixing the spurious local gates) can normally be omitted if swaps are performed in pairs.  Spectra are shown
for the pseudo-pure initial state $|0\rangle|+\rangle|0\rangle$ (an absorption signal on spin B) which is converted to $|0\rangle|-\rangle|0\rangle$ (an emission signal
on spin B) by the $S_z$ gate. Spectra in (c) show complete implementations of the Deutsch--Jozsa algorithm (analyzing functions from 2 bits to 1 bit) for the constant
function $f_{1111}$ and the balanced function $f_{0011}$ whose implementation requires a controlled-\textsc{not} operation between qubits A and C which are not directly
coupled.} \label{fig:intrablockExps}
\end{figure}

We therefore performed two families of experiments to demonstrate both classes of operation.  In each case we prepared spins in a pseudo-pure initial state by spatial
averaging \cite{Cory1997}.  For inter-block logic, we used the central spin as a buffer and implemented one- and two-qubit logic, employing these gates in a two-qubit
Deutsch algorithm \cite{Jones1998}, see Fig.~\ref{fig:interblockExps}. We emphasize that a chain {\em of any length} would use precisely the pulse sequences we have
experimentally realized on our three-spin device. Longer chains would simply employ extended periods of the regular inversion-generating global pulse sequence in order
to propagate qubits to the chain edge. In order to demonstrate intra-block logic we prepared our three-spin system with a single triple-qubit block as depicted in
Fig.~\ref{fig:newScheme}. Our experiments, shown in Fig.~\ref{fig:intrablockExps}, exhibit two crucial operations: a single qubit phase gate applied to the central
qubit, and a two-qubit gate between the outer qubits, which do not directly interact. The steps involved in the phase gate are shown in detail, while the use of the two
qubit gate is demonstrated in an instance of the Deutsch--Jozsa algorithm.

In conclusion, we have performed a series of experiments which constitute the first experimental demonstration of globally controlled quantum computation with
delocalized qubits.  Our demonstrations included the first quantum mirror to have been realized with a fully scalable procedure. We introduced a theoretical scheme for
computation with high density qubit storage, and realized all the key aspects of that scheme. Deutsch and Deutsch--Jozsa algorithms were implemented as example quantum
tasks.  Our new scheme minimizes the number of ancillary qubits so maximizing the computational power of a given spin chain. This work demonstrates the feasibility and power of the global control paradigm, and opens the
way to implementing QIP on a far wider range of systems than previously explored.

\section{Methods}
The inter-block gates previously described~\cite{twamley} do not require that each qubit be surrounded by buffers, but simply that a qubit have one neighbor in a buffer
state (see Supporting Materials).  Thus these techniques can be applied to the terminal qubits in a block of any length.  To apply gates to qubits \textit{inside} a
block requires these qubits to be permuted to the outside.  The permutation operations in Fig.~\ref{fig:newScheme} are performed using global control with the basic
operations $U_2=\CZ\,S_x^A\,\CZ$ and $U_1=H\,U_2\,H$; spurious local gates are most simply corrected by reversing the swap sequences after the desired local gate is
applied to a terminal spin. This network will work for any value of $M$, and moving selective pulses from the first spin to the last spin in a block allows gates to be
performed on the second last spin.  Qubits 2 and 3 can be swapped in the same way by replacing $U_2$ with $\CZ\,H\,U_2\,H\,\CZ$, and similarly for $U_1$. Successive
swap operations allow any single-qubit operations on \textit{any} qubit.  To implement two-qubit gates note that $U_2=\exp(-i\pi/4\,\sigma_x^A\sigma_z^B)$ is itself a
non-trivial two-qubit gate, and so in combination with single-qubit gates is universal.

The $\CZ$ operation can be implemented in an NMR system using a spin-echo pulse sequence, with global $z$-rotations either combined with Hadamard gates to give simpler
single-qubit gates or performed by frame rotations~\cite{Vandersypen2004}. Selective $z$-rotations may be achieved with composite Z-pulses~\cite{Vandersypen2004}. All
NMR experiments were performed on a Varian \mbox{Inova} 600\,MHz spectrometer, with ${}^{1}\text{H}$-decoupling applied during pulse sequences and acquisition to
simplify the spin system.  The NMR sample comprised 20\,mg of uniformly $^{13}\text{C}$-labeled alanine dissolved in 0.75\,ml of $\text{D}_2\text{O}$ at $25^\circ$C.
Strongly modulated composite pulses were designed to tolerate moderate rf inhomogeneity~\cite{Fortunato2002}.  Partial refocusing \cite{Bowdrey2005} was used to scale
down $J_{AB}$ to the same size as $J_{BC}$, and the small coupling $J_{AC}$ was neglected.  Initial states were prepared using conventional NMR techniques, but all
subsequent manipulations were performed using global control except where explicitly indicated.




\section{Acknowledgements}
JF is supported by a Helmore Award. SCB is supported by the Royal Society.  LX and JAJ thank the UK BBSRC and EPSRC for financial support.  We thank Jason Twamley for
helpful conversations.

Supplementary material is available for download at http://www.nano.org/globalPaper. 

\end{document}